\documentstyle[12pt,psfig]{article}
\begin{document}
\baselineskip 0.7cm

\newcommand{\gsim}{ \mathop{}_{\textstyle \sim}^{\textstyle >} }
\newcommand{\lsim}{ \mathop{}_{\textstyle \sim}^{\textstyle <} }
\newcommand{\vev}[1]{ \left\langle {#1} \right\rangle }
\newcommand{\lsp}{ \left ( }
\newcommand{\rsp}{ \right ) }
\newcommand{\lmp}{ \left \{ }
\newcommand{\rmp}{ \right \} }
\newcommand{\llp}{ \left [ }
\newcommand{\rlp}{ \right ] }
\newcommand{\labs}{ \left | }
\newcommand{\rabs}{ \right | }
\newcommand{\KEV}{ {\rm keV} }
\newcommand{\MEV}{ {\rm MeV} }
\newcommand{\GEV}{ {\rm GeV} }
\newcommand{\TEV}{ {\rm TeV} }
\newcommand{\mgut}{M_{GUT}}
\newcommand{\mint}{M_{I}}
\newcommand{\mgra}{M_{3/2}}
\newcommand{\mll}{m_{\tilde{l}L}^{2}}
\newcommand{\mdr}{m_{\tilde{d}R}^{2}}
\newcommand{\mllXX}[1]{m_{\tilde{l}L , {#1}}^{2}}
\newcommand{\mdrXX}[1]{m_{\tilde{d}R , {#1}}^{2}}
\newcommand{\mgy}{m_{G1}}
\newcommand{\mgl}{m_{G2}}
\newcommand{\mgc}{m_{G3}}
\newcommand{\nuR}{\nu_{R}}
\newcommand{\slL}{\tilde{l}_{L}}
\newcommand{\slLi}{\tilde{l}_{Li}}
\newcommand{\sdR}{\tilde{d}_{R}}
\newcommand{\sdRi}{\tilde{d}_{Ri}}
\newcommand{\e}{{\rm e}}

\begin{titlepage}

\begin{flushright}
KEK-TH-547,UT-793
\\
October, 1997
\end{flushright}

\vskip 0.35cm
\begin{center}
{\large \bf  Dimension-six Proton Decays \\
in the Modified Missing Doublet SU(5) Model}
\vskip 1.2cm
J.~Hisano$^{a)}$, Yasunori~Nomura$^{b)}$, and T.~Yanagida$^{b)}$

\vskip 0.4cm

a) {\it Theory Group, KEK, Tsukuba, Ibaraki 305, Japan}
\\
b) {\it Department of Physics, University of Tokyo, Tokyo 113, Japan}
\vskip 1.5cm

\abstract{Dimension-five operators for nucleon decays are suppressed
in the modified missing doublet (MMD) model in the supersymmetric SU(5)
grand unification.
We show that nonrenormalizable interactions decrease the unification scale
in the MMD model which increases the nucleon decay rate of 
dimension-six operators by a significant amount.
We find that the theoretical lower bound on the proton life time
$\tau(p \rightarrow \e^+ \pi^0)$ is within the observable range 
at SuperKamiokande.}

\end{center}
\end{titlepage}

%
%
%
%

The doublet-triplet splitting is one of the most serious 
problems in the supersymmetric (SUSY) grand unified theory (GUT).
Various interesting resolutions to this difficult problem have 
been proposed \cite{PLB115-380}\cite{PLB342-138}\cite{PLB-112}\cite{DW}\cite{PTP-75}\cite{PR-D53}. Among them the modified missing doublet
(MMD) SU(5) model \cite{PLB342-138} is the most attractive one, since it explains
very naturally the low value of the SU(3)$_C$ gauge coupling constant $\alpha_3(m_Z)= 0.118\pm 0.004$ 
\cite{alphas} by GUT-scale threshold 
corrections \cite{BMP}. Furthermore, the dangerous dimension-five
operators for nucleon decays \cite{NPB197-533} are strongly suppressed 
\cite{PLB342-138} owing to an additional chiral symmetry in the MMD model.

If the nucleon decay due to the dimension-five operators are outside the 
reach at SuperKamiokande, the dimension-six nucleon decays such as 
$p \rightarrow \e^+ \pi^0$ are clearly important to test the GUT. 
In this letter, we point out that introduction of nonrenormalizable
operators may decrease the GUT scale by a factor about 3 
for a reasonable parameter region in the
MMD SU(5) model. As a consequence the nucleon decays by the GUT gauge-boson
exchanges become $\sim 100$ times faster than the previous 
estimate \cite{NPB402-46}.
The present analysis, therefore, suggests that the theoretical lower bound 
for proton decay of the dominant channel $p \rightarrow \e^+ \pi^0$
is within the observable range at the SuperKamiokande detector.\footnote
{ Unfortunately, the MMD SU(5) model can not be excluded even if the nucleon
decays are not observed by the SuperKamiokande experiments.}

The original missing doublet (MD) model \cite{PLB115-380} in the SUSY SU(5) GUT
consists of the following chiral supermultiplets:
\begin{eqnarray}
&\psi_i ({\bf 10}),~\phi_i ({\bf 5}^\star),&
\nonumber\\
&H({\bf 5}),~\overline{H}({\bf 5}^\star), ~
\theta({\bf 50}),~\overline{\theta}({\bf 50}^\star),~
\Sigma({\bf 75}),&
\nonumber
\end{eqnarray}
where $i(=1-3)$ represents family index. This original MD model
is incomplete, since the SU(5) gauge coupling constant 
$\alpha_5(\equiv g^2_5/4 \pi)$
blows up below the gravitational scale $M_G \simeq 2.4 \times 10^{18}
$GeV because of the large representations for $\theta$, $\overline{\theta}$,
and $\Sigma$. One may avoid this unwanted situation by taking a mass for 
 $\theta$ and $\overline{\theta}$ at the gravitational scale. However,
in this case the colored Higgs mass $M_{H_c}$ becomes lower than
$\sim 10^{15}$ GeV which leads to too rapid nucleon decays by
dimension-five operators \cite{NPB402-46}. 

The MMD SU(5) model \cite{PLB342-138} solves the above problem by imposing 
an additional chiral (Peccei-Quinn) symmetry. To incorporate 
the Peccei-Quinn symmetry in the original MD model, we introduce
a new set of chiral supermultiplets, 
\begin{eqnarray}
&H^\prime({\bf 5}),~\overline{H}^\prime({\bf 5}^\star),~ 
\theta^\prime({\bf 50}),~\overline{\theta}^\prime({\bf 50}^\star).&
\nonumber
\end{eqnarray}
We, then, assume the U(1)$_{\rm PQ}$ charges for each multiplets as
\begin{eqnarray}
\psi_i({\bf 10}) \rightarrow \e^{i\alpha/2} \psi_i({\bf 10}), &&
\phi_i({\bf 5}^\star) \rightarrow \e^{i\beta/2} \phi_i({\bf 5}^\star),
\nonumber \\
H ({\bf 5})\rightarrow \e^{-i\alpha} H ({\bf 5}), &&
\overline{H}({\bf 5^\star}) \rightarrow \e^{-i\frac{\alpha+\beta}{2}} \overline{H} ({\bf 5}^\star),
\nonumber \\
H^\prime ({\bf 5})\rightarrow \e^{i\frac{\alpha+\beta}{2}} H^\prime ({\bf 5}), &&
\overline{H}^\prime ({\bf 5^\star}) \rightarrow \e^{i\alpha} \overline{H}^\prime ({\bf 5}^\star),
\nonumber \\
\theta ({\bf 50}) \rightarrow \e^{i\alpha} \theta ({\bf 50}),&&
\overline{\theta} ({\bf 50}^\star)\rightarrow \e^{i\frac{\alpha+\beta}{2}} 
\overline{\theta} ({\bf 50}^\star),
\nonumber \\
\theta^\prime ({\bf 50}) \rightarrow \e^{-i\frac{\alpha+\beta}{2}} \theta^\prime ({\bf 50}),&&
\overline{\theta}^\prime ({\bf 50}^\star) \rightarrow \e^{-i\alpha} 
\overline{\theta}^\prime ({\bf 50}^\star),
\nonumber \\
\Sigma ({\bf 75})\rightarrow \Sigma ({\bf 75}),&&
\label{PQ_charge}
\end{eqnarray}
with $3\alpha+\beta \ne 0$. 

We take masses of $\theta$, $\overline{\theta}^\prime$, and 
$\theta^\prime$, $\overline{\theta}$ at the gravitational 
scale $M_G$ to maintain the perturbative description of the SUSY
GUT. Then, we have only two pairs of Higgs multiplets, $H$, 
$\overline{H}^\prime$, and $H^\prime$, $\overline{H}$, and one
Higgs $\Sigma$ in addition to three families of quark and lepton
multiplets below the scale $M_G$.

The 75-dimensional Higgs $\Sigma$ has the following vacuum expectation
value causing the desired SU(5)$\rightarrow$SU(3)$_C\times$SU(2)$_L\times$U(1)$_Y$ breaking,
\begin{eqnarray}
\langle \Sigma \rangle_{(\gamma\delta)}^{(\alpha\beta)} 
&=& \frac12 \left\{
\delta_\gamma^\alpha \delta_\delta^\beta -
\delta_\delta^\alpha \delta_\gamma^\beta
\right\} V_{\Sigma},
\nonumber\\
\langle \Sigma \rangle_{(cd)}^{(ab)} 
&=& \frac32 \left\{
\delta_c^a \delta_d^b -
\delta_d^a \delta_c^b
\right\} V_{\Sigma},
\\
\langle \Sigma \rangle_{(b \beta)}^{(a \alpha)} 
&=& -\frac12 \left\{
\delta_b^a \delta_\beta^\alpha
\right\} V_{\Sigma},
\nonumber
\end{eqnarray}
where $\alpha$, $\beta$, $\cdots$ are the SU(3)$_C$ indices and 
$a$, $b$, $\cdots$ the SU(2)$_L$ indices. Integration of the 
heavy fields, $\theta$, $\overline{\theta}^\prime$, and 
$\theta^\prime$, $\overline{\theta}$, generates 
Peccei-Quinn invariant masses of the colored
Higgs multiplets as 
\begin{eqnarray}
  M_{H_c} H_c^\alpha \overline{H}_{c\alpha}^{\prime} 
+ M_{\overline{H}_c} H_c^{\prime \alpha} \overline{H}_{c\alpha}
\end{eqnarray}
with
\begin{eqnarray}
\label{coloredmasses}
M_{H_c} \simeq 48 G_H G_{\overline{H}}^{'} \frac{V_{\Sigma}^2}{M_G},~~~
M_{\overline{H}_c} \simeq 48 G_{\overline{H}}G_H^{'} \frac{V_{\Sigma}^2}{M_G}.
\end{eqnarray}
Here, the coupling constants $G_H$, $G_{\overline{H}}$, $G_H^{'}$,
and $G_{\overline{H}}^{'}$ are expected to be $O(1)$ 
(see Ref.~\cite{PLB342-138} for their definition). 
Notice that dimension-five operators for nucleon decays are 
completely suppressed as long as the Peccei-Quinn symmetry 
is unbroken.

The two pairs of Higgs doublets, $H_f$, 
$\overline{H}_f$, and  $H_f^{\prime}$, $\overline{H}_f^{\prime}$, on the 
other hand, remain massless. They acquire masses through the Peccei-Quinn
symmetry breaking. As shown in Ref.~\cite{PLB342-138} we choose 
U(1)$_{\rm PQ}$ charges for 
the Peccei-Quinn symmetry breaking fields so that only 
one pair of Higgs doublets   $H_f^{\prime}$ and $\overline{H}_f^{\prime}$
has a mass $M_{H_f^\prime}$ of the order of the Peccei-Quinn scale 
$\sim (10^{10}-10^{12})$GeV.
Thus, the model is nothing but the SUSY standard model 
below the Peccei-Quinn scale.
The Peccei-Quinn symmetry breaking generates the dimension-five operators
which are, however, proportional to 
$M_{H_f^\prime}/(M_{H_c}M_{\overline{H}_c})$.

We are now ready to discuss renormalization group (RG) equations for 
the SU(3)$_C\times$SU(2)$_L\times$U(1)$_Y$ gauge coupling constants 
$\alpha_3$, $\alpha_2$, and $\alpha_1$ in the standard model.
It has been pointed out in Ref.~\cite{PRL-69} that the GUT scale and the colored Higgs 
mass are determined independently by using the gauge coupling constants
at the electroweak scale. 
For the case of the MMD model we have \cite{PLB342-138}
at the one-loop level
\begin{eqnarray}
\label{MHC}
(3 \alpha_2^{-1} - 2 \alpha_3^{-1} - \alpha_1^{-1}) (m_Z)
        &=& \frac{1}{2\pi} \Bigg\{ 
                \frac{12}{5} \, \ln \frac{M_{H_c}^{eff}}{m_Z}
                - 2 \, \ln \frac{m_{\rm SUSY}}{m_Z} \nonumber\\
&&               - \frac{12}{5} \, \ln (1.7\times 10^4)
\Bigg\},\\
\label{MVMSIGMA}
(5 \alpha_1^{-1} - 3 \alpha_2^{-1} - 2 \alpha_3^{-1}) (m_Z)
        &=& \frac{1}{2\pi} \Bigg\{
                12 \, \ln \frac{M_V^2 M_\Sigma}{m_Z^3}
                + 8 \ln \frac{m_{\rm SUSY}}{m_Z}  \nonumber\\ 
&&               + 36 \, \ln (1.4)
\Bigg\}.
\end{eqnarray}
Here, the effective mass for the colored Higgs 
$M_{H_c}^{eff} = M_{H_c}M_{\overline{H}_c}/M_{H_f^\prime}$.
Notice that the last values in Eqs.~(\ref{MHC}) and (\ref{MVMSIGMA})
come from the mass splitting among the components in $\Sigma({\bf 75})$.
$M_V$ is the mass for the GUT gauge bosons 
($M_V =2 \sqrt{6}g_5 V_{\Sigma}$),\footnote{
We find a mistake in a form of $M_V$ in Ref.~\cite{PLB342-138}.
}
and $M_{\Sigma}$ the mass of the heaviest component of $\Sigma({\bf 75})$
($M_{\Sigma} = (10/3) \lambda_{75} V_{\Sigma}$).
The definition of Yukawa coupling constant $\lambda_{75}$ 
is given in Ref.~\cite{PLB342-138}.

To perform a quantitative analysis we use two-loop RG 
equations between the electroweak and the GUT scale. For 
GUT scale threshold corrections we use the one-loop result. We use 
the mass spectrum derived from the minimum supergravity instead of the common
mass $m_{\rm SUSY}$ for superparticles to calculate one-loop threshold 
corrections at the SUSY-breaking scale. Using the experimental data
$\alpha_{em}^{-1}(m_Z) = 127.90\pm 0.09$, 
$\sin^2\theta_W(m_Z) =0.2314 \pm 0.0004$ \cite{weinberg_angle} 
and $\alpha_3(m_Z) = 0.118 \pm 0.004$ \cite{alphas}, we obtain the GUT scale
as
\begin{eqnarray}
9.5 \times 10^{15}~\GEV \leq 
&(M_V^2 M_\Sigma)^{1/3}
& \leq 2.3 \times 10^{16}~\GEV.
\label{GUTscale}
\end{eqnarray}

The dimension-six operators for nucleon decays depend on $M_V$, and 
hence Eq.~(\ref{GUTscale}) is not sufficient to calculate the decay 
rate. Fortunately, the lower bound on $M_V$ is given by the following
consideration. The upper bound $M_{\Sigma}$ is determined by requiring
that the Yukawa coupling $\lambda_{75}$ never blows up below the gravitational
scale. By solving the one-loop RG equation for $\lambda_{75}$
we find\footnote{
When $\lambda_{75}$ is much larger than $g_5$ at the gravitational scale, 
it rapidly closes to the quasi-infrared fixed point at the GUT scale, 
and on the fixed point $M_{\Sigma}/M_V = 1.28$.
}
\begin{equation}
M_{\Sigma} \le 1.3 M_V,
\label{infrared}
\end{equation}
which leads to together with Eq.~(\ref{GUTscale})
\begin{equation}
M_{V} \ge  8.7 \times 10^{15}\GEV.
\end{equation}
This gives the lower bound for the proton life time as 
\begin{eqnarray}
\tau(p \rightarrow \e^+ \pi^0) 
&=&
2.9 \times 10^{35} \times \left(\frac{M_V}{10^{16}\GEV}\right)^4
\left(\frac{0.0058\GEV^3}{\alpha}\right)^2 {\rm years}
\nonumber\\
&\ge&  1.6 \times 10^{35} {\rm years}.
\end{eqnarray}
Here, $\alpha$ is the hadron matrix element\footnote
{See Ref.~\cite{NPB402-46} for the definition of $\alpha$.}
and we have used a lattice value $\alpha = 0.0058\GEV^3$ \cite{NPB312}.

One may consider that threshold corrections from some additional
particles at the GUT scale could decrease the GUT scale. However,
if one introduces additional multiplets at the GUT scale, the SU(5)
gauge coupling constant tends to blow up below the gravitational 
scale.
Simple choices may be pairs of ${\bf 5}+{\bf 5}^\star$ and/or 
pairs of ${\bf 10}+{\bf 10}^\star$. 
We easily see from Eq.~(\ref{MVMSIGMA}) that the pairs of 
${\bf 5}+{\bf 5}^\star$ do not change the GUT scale $(M_V^2 M_{\Sigma})^{1/3}$.
As for ${\bf 10}+{\bf 10}^\star$ we have checked that any significant change
of the GUT scale is not obtained by their threshold corrections for a natural
range of mass parameters.

We are now led to consider nonrenormalizable interactions suppressed by
$1/M_G$. The most dominant operator contributing to the gauge 
coupling constants $\alpha_3$, $\alpha_2$, and $\alpha_1$ is
\begin{equation}
\int d^2 \theta \frac{\delta}{M_G} {\cal W}_B^A \cdot 
{\cal W}_D^C ~\Sigma^{(BD)}_{(AC)} + {\rm h.c.},
\label{NRO}
\end{equation}
where $A$, $B$, $\cdots$ are SU(5) indices.
Here, ${\cal W}$ is the SU(5) field-strength superfields 
whose kinetic term is given by 
\begin{equation}
\int d^2 \theta \frac{1}{8 g^2_5} {\cal W}_B^A \cdot {\cal W}_A^B
 + {\rm h.c.}.
\end{equation}
The above nonrenormalizable term yields corrections to 
$\alpha_i(i=1-3)$ at the GUT scale as 
\begin{eqnarray}
\alpha_3^{-1} &=& \alpha_5^{-1} - 4\left(4 \pi \delta \frac{V_{\Sigma}}{M_G}\right),  
\nonumber\\
\alpha_2^{-1} &=& \alpha_5^{-1} - 12\left(4 \pi \delta \frac{V_{\Sigma}}{M_G}\right),  
\\
\alpha_1^{-1} &=& \alpha_5^{-1} + 20\left(4 \pi \delta\frac{V_{\Sigma}}{M_G}\right).  
\nonumber
\end{eqnarray}
Then, Eqs.~(\ref{MHC}) and (\ref{MVMSIGMA}) are modified as 
\begin{eqnarray}
\label{MHC1}
(3 \alpha_2^{-1} - 2 \alpha_3^{-1} - \alpha_1^{-1}) (m_Z)
        &=& \frac{1}{2\pi} \Bigg\{ 
                \frac{12}{5} \, \ln \frac{M_{H_c}^{eff}}{m_Z}
                - 2 \, \ln \frac{m_{\rm SUSY}}{m_Z} \nonumber\\
&&               - 48\, (8 \pi^2 \delta \frac{V_{\Sigma}}{M_G})
                  - \frac{12}{5} \, \ln (1.7\times 10^4)
\Bigg\},\\
\label{MVMSIGMA1}
(5 \alpha_1^{-1} - 3 \alpha_2^{-1} - 2 \alpha_3^{-1}) (m_Z)
        &=& \frac{1}{2\pi} \Bigg\{
                12 \, \ln \frac{M_V^2 M_\Sigma}{m_Z^3}
                + 8\, \ln \frac{m_{\rm SUSY}}{m_Z}  \nonumber\\ 
&&               + 144\, (8 \pi^2 \delta \frac{V_{\Sigma}}{M_G})
                 + 36 \, \ln (1.4)
\Bigg\}.
\end{eqnarray}
We see that  the GUT scale $(M_V^2 M_{\Sigma})^{1/3}$
decreases easily by a factor $\sim3$ for 
$\delta V_{\Sigma} /M_G=0.0035$.\footnote{
Introduction of a similar nonrenormalizable operator 
to Eq.~(\ref{NRO}) suppressed by $1/M_G$ 
does not change the GUT scale in the minimal SUSY SU(5) GUT \cite{PRL70-2676}. 
}
As a consequence the effective mass for the colored Higgs multiplets
$M_{H_c}^{eff}$ increases by almost 240 times compared with the values 
for $\delta=0$. This is rather welcome to suppress the dimension-five 
operators for nucleon decays since they are proportional to 
$1/M_{H_c}^{eff}$. Notice that the presence of the 
nonrenormalizable operator in Eq.~(\ref{NRO}) lowers further the 
value of $\alpha_3(m_Z)$  for $\delta>0$. This over reduction of 
$\alpha_3(m_Z)$ is, however, compensated by raising the $M_{H_c}^{eff}$.

\begin{figure}
  \centerline{\psfig{file=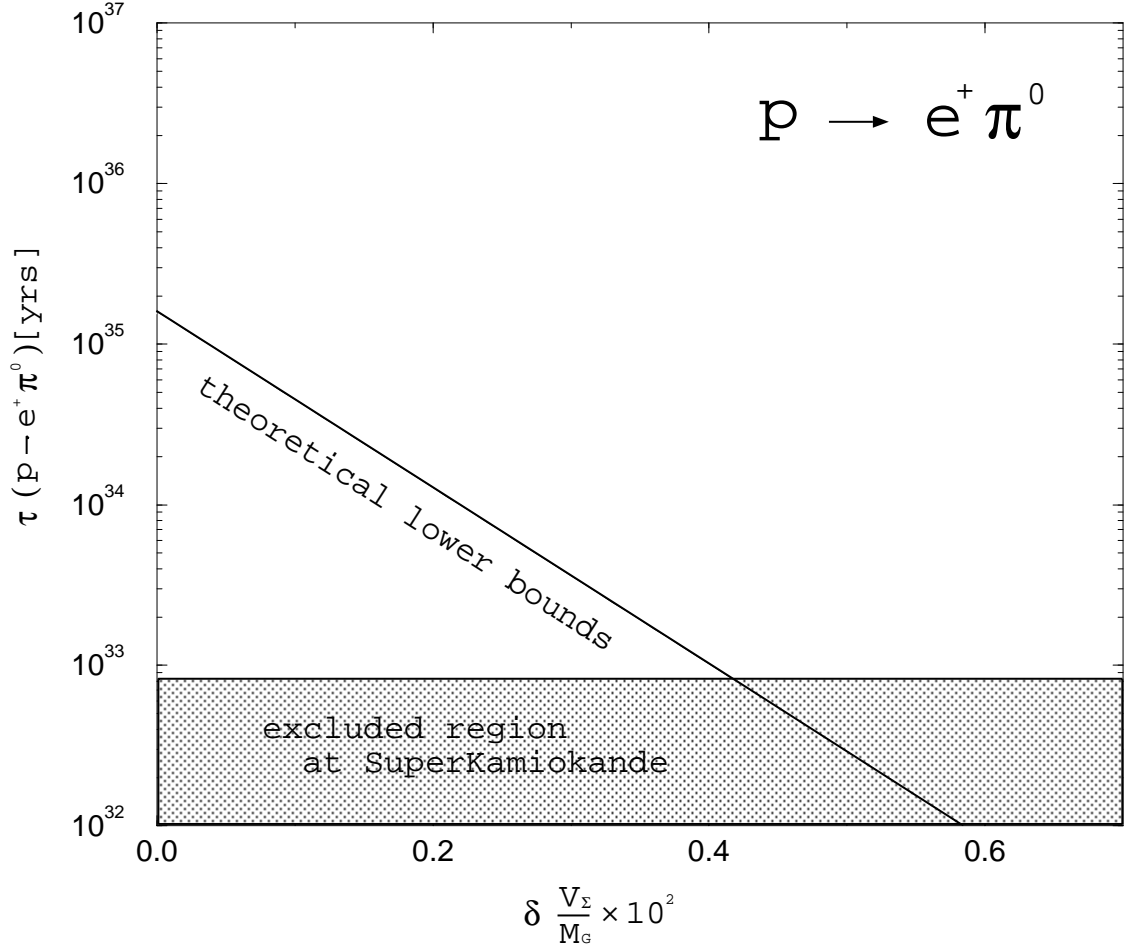,width=15cm}}
\caption{The theoretical lower bounds on the proton lifetime 
  $\tau(p \rightarrow \e^+ \pi^0)$ as a function of the parameter 
  $(\delta V_{\Sigma}/M_{G})\times 10^{2}$.
  Here, we have assumed a lattice value of the hadron matrix element 
  $\alpha = 0.0058\GEV^3$ \cite{NPB312}.
  The shaded area is excluded by the recent SuperKamiokande experiments
  ( $\tau(p \rightarrow \e^+ \pi^0) > 7.9 \times 10^{32} {\rm years}$
  \cite{kamioka} ).}
\label{lifetime_figure}
\end{figure}

In conclusion, we have shown in this letter that introduction of 
the nonrenormalizable interaction may decrease the GUT scale 
by a factor $\sim3$ in the MMD SU(5) model \cite{PLB342-138}. 
In Fig.~{\ref{lifetime_figure}}, 
we show the theoretical lower bounds on the proton lifetime 
$\tau(p \rightarrow \e^+ \pi^0)$.
We see that they lie within the observable range at 
SuperKamiokande for a reasonable parameter region 
$\delta \sim O(1)$.
A recent report \cite{kamioka} from SuperKamiokande shows that 
one event of the mode $\e^+ \pi^0$ 
still survives the cut criterion for backgrounds. If this is 
indeed a real signal for $p \rightarrow \e^+ \pi^0$ decay, we may be led
to a serious consideration of physics at the gravitational scale.
\newpage

%
%
\newcommand{\Journal}[4]{{\sl #1} {\bf #2} {(#3)} {#4}}
\newcommand{\PL}{\sl Phys. Lett.}
\newcommand{\PR}{\sl Phys. Rev.}
\newcommand{\PRL}{\sl Phys. Rev. Lett.}
\newcommand{\NP}{\sl Nucl. Phys.}
\newcommand{\ZP}{\sl Z. Phys.}
\newcommand{\PTP}{\sl Prog. Theor. Phys.}
\newcommand{\NC}{\sl Nuovo Cimento}

\end{document}